\documentclass[sigconf]{acmart}

\AtBeginDocument{%
  \providecommand\BibTeX{{%
    \normalfont B\kern-0.5em{\scshape i\kern-0.25em b}\kern-0.8em\TeX}}}

\usepackage[utf8]{inputenc} 
\usepackage[T1]{fontenc}    
\usepackage{hyperref}       
\usepackage{url}            
\usepackage{booktabs}       
\usepackage{amsfonts}       
\usepackage{nicefrac}       
\usepackage{microtype}      

\usepackage{graphicx}

\usepackage{amsmath,bbm,amsthm}
\usepackage{epsfig} 
\usepackage{multicol}
\usepackage{multirow}
\usepackage{color}
\usepackage{mathrsfs}
\usepackage{xparse}
\usepackage{algorithm}
\usepackage{algorithmicx}
\usepackage{algpseudocode}
\usepackage{wrapfig}
\usepackage[font=small,labelfont=bf,tableposition=top]{caption}
\usepackage[font=footnotesize]{subcaption}

\usepackage{mathtools}

\usepackage{optidef}
\usepackage{gensymb}
\usepackage{acronym}
\acrodef{ML}{Machine Learning}
\acrodef{AI}{Artificial Intelligence}
\acrodef{TTA}{Test-time Augmentation}
\acrodef{SL}{Severity Level}
\acrodef{i.d.}{in-distribution}
\acrodef{o.o.d.}{out-of-distribution}
\acrodef{DL}{Deep Learning}
\acrodef{KL}{Kullback-Leibler}
\acrodef{DSE}{Design Space Exploration}
\acrodef{FPR}{False Positive Rate}
\acrodef{FNR}{False Negative Rate}
\acrodef{TPR}{True Positive Rate}
\acrodef{TNR}{True Negative Rate}
\acrodef{FNP}{False Negative Precision}
\acrodef{AUC}{Area Under Curve}
\acrodef{FDD}{Fault Detection and Diagnosis}
\acrodef{CNN}{Convolutional Neural Network}
\acrodef{GAN}{Generative Adversarial Network}
\acrodef{MSE}{Mean Squared Error}
\acrodef{SRR}{Super-Resolution Reconstruction}
\acrodef{AMI}{Advanced Metering Infrastructure}
\acrodef{WPE}{Window Peak Error}

\setcopyright{rightsretained}

\begin{document}

\title{Poster Abstract: Super-Resolution Reconstruction of\\Interval Energy Data}


\author{Jieyi Lu}
\affiliation{%
  \institution{Yale School of the Environment}
  \streetaddress{195 Prospect St}
  \city{New Haven}
  \country{CT~06511}}
\email{jieyi.lu@yale.edu}

\author{Baihong Jin}
\affiliation{%
  \institution{EECS Department, UC Berkeley}
  \streetaddress{Cory Hall}
  \city{Berkeley}
  \country{CA~94720}}
\email{bjin@eecs.berkeley.edu}

\renewcommand{\shortauthors}{Jieyi Lu and Baihong Jin}

\begin{abstract}
    High-resolution data are desired in many data-driven applications; however, in many cases only data whose resolution is lower than expected are available due to various reasons. It is then a challenge how to obtain as much useful information as possible from the low-resolution data. In this paper, we target \textit{interval energy data} collected by \ac{AMI}, and propose a \ac{SRR} approach to upsample low-resolution (hourly) interval data into higher-resolution (15-minute) data using deep learning. Our preliminary results show that the proposed \ac{SRR} approaches can achieve much improved performance compared to the baseline model.
\end{abstract}

\copyrightyear{2020}
\acmYear{2020}
\acmConference[BuildSys '20]{The 7th ACM International Conference on Systems for Energy-Efficient Buildings, Cities, and Transportation}{November 18--20, 2020}{Virtual Event, Japan}
\acmBooktitle{The 7th ACM International Conference on Systems for Energy-Efficient Buildings, Cities, and Transportation (BuildSys '20), November 18--20, 2020, Virtual Event, Japan}\acmDOI{10.1145/3408308.3431115}
\acmISBN{978-1-4503-8061-4/20/11}


\maketitle

\vspace{-2mm}
\section{Introduction}
\acresetall
Interval energy data (a.k.a.~\textit{interval data}) are fine-grained records of energy consumption captured by \ac{AMI}; the readings are made at regular time intervals. Each measurement represents the total energy consumption (in kWh) during the corresponding time interval.

Interval data contains rich information on energy consumption of different users and can be useful in many cases. First, interval data can identify energy use peaks and trends, which can be used as a diagnostic tool for anomaly detection~\cite{parker2015metering}. Second, it can inform and evaluate energy efficiency policies. Researchers evaluated the performance of policies, such as electricity subsidies and peak electricity pricing, by analyzing the variations in electricity consumption informed by interval data~\cite{jessoe2014knowledge, sherwin2020characterizing}.


Scanning through the literature, we discover that interval data of various resolutions have been used for control and analysis, where 15-minute and 60-minute (hourly) interval data being the most common formats for residential users in the U.S. However, accessing finer interval data is not easy. Permission from customers and web-based systems for downloading data are required~\cite{eia2015assessment}. Although the \textit{resolution mismatch} problem can always be mitigated by downsampling or upsampling, it is still unknown whether such simple solutions will incur interoperability issues when we apply models trained on one resolution to data of another resolution.

To address the above-mentioned issue, we propose to use a deep-learning-based \ac{SRR} technique for obtaining high-quality upsampling results from low-resolution interval data to improve the quality of low-resolution data and recover, to a possible extent, the contents in the original high-resolution signal. In this paper, we focus mainly on interval data whose sampling rate is in the sub-hourly domain, specifically, how to design a deep learning model to upsample hourly interval data into 15-minute interval data.


\begin{figure}[tb]
    \centering
      \begin{subfigure}[t]{0.48\columnwidth}
        \centering
        \includegraphics[height=3cm]{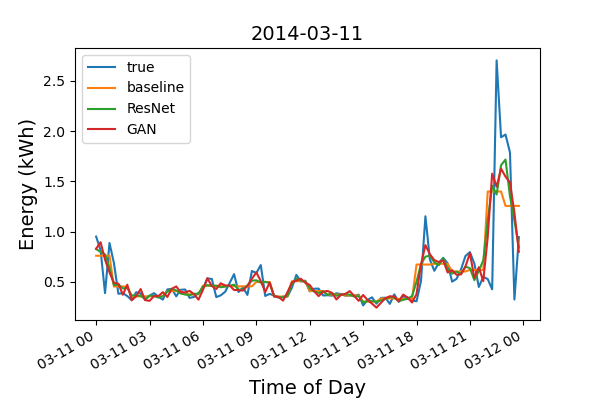}
        \caption{Household No.9612 (California)}
      \end{subfigure}
      \begin{subfigure}[t]{0.48\columnwidth}
        \centering
        \includegraphics[height=3cm]{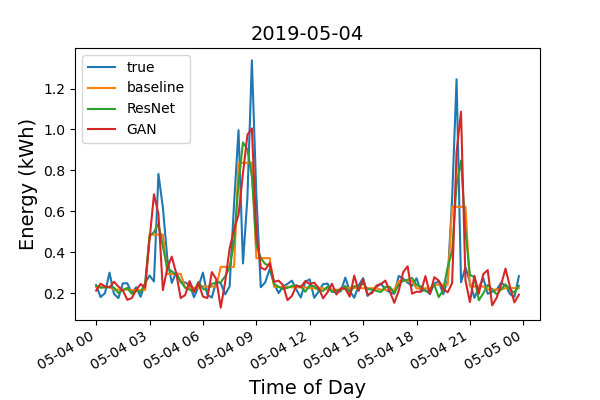}
        \caption{Household No.2096 (New York)}
      \end{subfigure}
      
      \begin{subfigure}[t]{0.48\columnwidth}
        \centering
        \includegraphics[height=3cm]{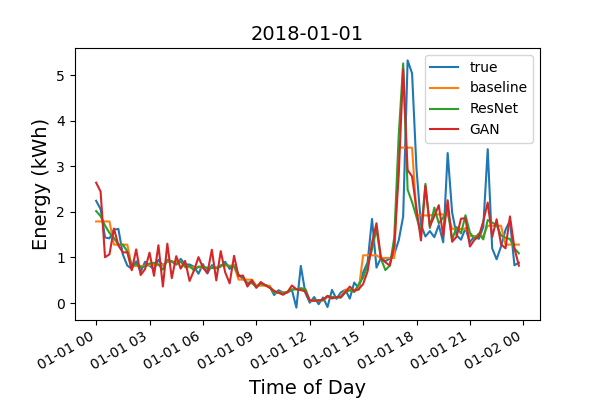}
        \caption{Household No.2096 (Texas)}
      \end{subfigure}
    \vspace{-3mm}
    \caption{\ac{SRR} results on three households from the test set, one from each state: California, New York, and Texas.}
    \label{fig:selected-results}
    \vspace{-5mm}
\end{figure}

\vspace{-2mm}
\section{Problem Formulation}

We consider an integer factor $S$ for upsampling each interval of time length $\tau$ into $S$ intervals each with length $\nicefrac{S}{\tau}$. Let us denote by $\{s_k\}=s_1s_2\ldots s_k\ldots s_K$ a low-resolution signal of $K$ $\tau$-intervals, and by 
$\{r_k\}=(r_1^1r_1^2\ldots r_1^S)(r_2^1r_2^2\ldots r_2^S)\ldots(r_k^1 r_k^2 \ldots r_k^S)\ldots(r_K^1 r_K^2 \ldots r_K^S)$ the corresponding high-resolution signal. Each entry in sequences $\{s_k\}$ and $\{r_k\}$ represents the energy consumption during the corresponding interval. 
To achieve \ac{SRR}, we aim to learn a function $f: \mathbb{R}^T\rightarrow\mathbb{R}^{S\cdot T}$ that maps $\{s_k\}$ to $\{r_k\}$ while respecting the following energy constraint that connects the two signals of different resolutions,
\vspace{-5mm}
\begin{align}\label{eqn:interval-energy}
    s_k = r_k^1 + r_k^2 + \ldots + r_k^S,~\forall k.
\end{align}
We evaluate the reconstruction quality from two different angles. First, as in many other regression tasks, we measure the \ac{MSE} between  $\{r_k\}$ and $\{\hat{r}_k\}$. In addition, we introduce another error metric, \ac{WPE}, to measure how accurately the reconstructed signal captures the peak (maximum) value of each $W$-hour windowed signal $w$ (in terms of their absolute difference). The \ac{WPE} is calculated as the mean error across all sliding windows of size $W$.

\vspace{-2mm}
\section{Methodology}

\vspace{-1.5mm}
\paragraph{Baseline Method}
We use a \textit{piece-wise constant interpolation} approach as the baseline method for comparison purpose. Although there are other off-the-shelf interpolation methods available (e.g., linear or bicubic interpolation), most of these methods do not automatically respect the interval energy constraint~\eqref{eqn:interval-energy} without modification. The piece-wise constant interpolation baseline can be easy to implement. For each interval $s_k$ in the low-resolution signal, the corresponding high-resolution reconstruction signal $\{\hat{r}_k\}$ for that interval can be calculated by simply letting $\hat{r}_k^1 = \hat{r}_k^2 = \ldots = \hat{r}_k^S = \nicefrac{s_k}{S}$.

\vspace{-1.5mm}
\paragraph{\ac{CNN} Method}
Our deep learning approaches for upsampling energy interval data are inspired by the popular image \ac{SRR} approach~\cite{ledig2017photo}. Instead of the 2D, multi-channel data, we are dealing with 1D, single-channel interval data as the model input in our scenario. Let us introduce a set of $S$ \textit{allocation factors} $\alpha_1,\alpha_2,\ldots,\alpha_S$ for each $s_k$ such that $\sum_{j=1}^S \alpha_k^j = 1$ and $\hat{r}_k^j =\alpha_k^j r_k$ for every $j$.
The values of $\alpha_k^j$ represent an \textit{allocation} of energy across finer-resolution time intervals, and they should sum up to one in order to respect the interval energy constraint~\eqref{eqn:interval-energy}. Note that $\alpha_k^j$ can take negative values to represent a negative net energy consumption, for example, when the household is exporting excessive solar generation to the grid.
To perform \ac{SRR} for low-resolution interval data, we can design a \ac{CNN} model to predict the allocation factors $\alpha_k^j$'s for each time interval $k$; the \ac{CNN} model can be understood as a \textit{filter} that transforms the signal into a set of allocation factors that respects the constraint $\sum_{j=1}^S \alpha_k^j = 1$.

\vspace{-1.5mm}
\paragraph{\ac{GAN} Method}
The \ac{CNN} method can give a very small \ac{MSE} as directed by the training goal. The \ac{MSE}, however, may not be a good measure for reconstruction quality because the baseline method is likely to achieve a small \ac{MSE} already. Thus, a small difference in \ac{MSE} between the baseline and the \ac{CNN} method does not necessarily indicate a good performance. To overcome this difficulty, we resort to a \ac{GAN}-based approach~\cite{ledig2017photo}. A \ac{GAN} model~\cite{goodfellow2014generative} consists of two parts, a \textit{generator} and a \textit{discriminator}. The generator model aims to produce accurate \ac{SRR} interval data from low-resolution ones, and the discriminator aims to differentiate the generated data and the real interval data. By training the generator and the discriminator jointly in an adversarial fashion, they can improve each other. After training, the resulting generator network can be used as the \ac{SRR} model.

\vspace{-2mm}
\section{Experimental Study}
We tested our proposed method on the Pecan Street dataset~\cite{Smith2009ThePS}, using the publicly available data from 73 households across three states in the U.S., New York (NY), California (CA), and Texas (TX). We used 15 CA households and 18 NY households for training, and used the rest for testing. 


\begin{table}[tb]
\centering
\caption{Experimental Results}
\label{tab:srr-loss}
\vspace{-3mm}
\resizebox{\columnwidth}{!}{%
\begin{tabular}{ccccccc}
\hline
\multicolumn{1}{l}{} & \multicolumn{2}{c}{California (test)} & \multicolumn{2}{c}{New York (test)} & \multicolumn{2}{c}{Texas} \\
\multicolumn{1}{l}{} & MSE            & mean WPE       & MSE            & mean WPE       & MSE            & Peak           \\ \hline
Baseline             & \textbf{0.058} & 0.231          & \textbf{0.325} & 0.491          & \textbf{0.429} & 0.669          \\
CNN                  & 0.073          & \textbf{0.172} & 0.345          & \textbf{0.383} & 0.569          & \textbf{0.511} \\
GAN                  & 0.109          & 0.192          & 0.460          & 0.431          & 0.555          & 0.743          \\ \hline
\end{tabular}%
}
\vspace{-5mm}
\end{table}


The performance of the three methods (baseline, \ac{CNN} and \ac{GAN}) in terms of \ac{MSE} and \ac{WPE} is shown in Table~\ref{tab:srr-loss}. From the results we can see that both \ac{CNN} and \ac{GAN} approaches lead to slightly increased \ac{MSE} on test set data, but achieve much better performance in terms of the \ac{WPE} metric, which indicates that our proposed methods can more faithfully recover the peaks in the original, high-resolution data. The displayed \ac{WPE} results are for window size $W=3$; we also evaluate the performance for other window sizes and the results are similar.

We also show a few \ac{SRR} results in Fig.~\ref{fig:selected-results}. As can be seen from the plots, the \ac{GAN}-based method excels at capturing the peak energy consumption events, despite sometimes missing the accurate occurrences of the peaks. The misalignment of the peaks will result in increased errors under the \ac{MSE} metric, but will not be penalized by our proposed peak consumption error metric if the temporal misalignment is not significant. The \ac{CNN} method can also capture some of the peak events; however, the resulting load profile is overly smooth and cannot predict well the magnitude of the peaks. This demonstrates the benefits of incorporating a discriminator along with the \ac{CNN}-based generator in the \ac{GAN}-based approach: the discriminator encourages the generator to produce more authentic load profiles besides lowering the \ac{MSE}.

\vspace{-2mm}
\section{Summary}
The presented results have several implications. First, we have shown that by using deep learning approaches we can recover to a good extent the important information of high-resolution (15-minute) interval data that are missing from lower-resolution (hourly) data. The reconstructed high-resolution data can be potentially useful for household energy management applications such as \textit{peak shaving} and \textit{battery dispatch}. In addition, the findings motivate us to rethink potential privacy issues associated with low-resolution interval data.  For future work, we plan to further explore the design space for better performing models.

\bibliographystyle{acm}
\bibliography{refs}

\end{document}